# A study on baryons spectroscopy using digamma-function as interacting potential

L. I. Abou-Salem

*Physics department, faculty of Science, Benha University, Benha, Egypt*

Email: loutfy.abousalem@fsc.bu.edu.eg

## Abstract

In the present work, the interacting potential form between quarks inside a baryon is taken to be proportional to the digamma-function. Using the Jacobi-coordinates the three body wave equation is solved to calculate the different states of the considered baryons. The present model contains only two adjustable parameters in addition to the quark masses. Our theoretical calculations are compared to the available experimental data and Cornelll potential calculations. The present potential model calculations show a satisfied agreement in case of investigated the resonance masses of the considered baryons. One can conclude that; the interaction between the quark constituents of baryon systems could be described adequately by using the considered simple potential form.

## Introduction

The hadrons spectroscopy is very important to study its structures and the nature of the interacting forces between its constituents. In the previous works [1-5], the heavy mesons spectroscopy is studied by using different potential models. Many authors have been studied the baryon spectroscopy. An analytical solution was provided in case of harmonic and inharmonic potentials for a system consist of three identical particles [6-7]. Some authors have been used the Cornell potential form in case of studding the resonance states of N and Δ baryons [8-13]. The harmonic oscillator potential has been used [14] to study the three interaction between three identical particles. In the present work we used the digamma-function as the interacting potential between constituents quarks of baryon systems. The three body wave equation is solved; numerically; by using the Jacobi method to calculate the resonance state masses of N, Δ, Λ and Σ baryons. These baryons consist of identical or non-identical quarks. Our theoretical results will be investigated in the results and discussion section.

**The Used Model**

Schrodinger equation for a system consists of three particles is given by:

$$\left[-\frac{\hbar^2}{2m_1}\nabla_{r_1}^2 - \frac{\hbar^2}{2m_2}\nabla_{r_2}^2 - \frac{\hbar^2}{2m_3}\nabla_{r_3}^2 + V_{12}(r_{12}) + V_{23}(r_{23}) + V_{31}(r_{31})\right]\Psi(r_1, r_2, r_3) = E\Psi(r_1, r_2, r_3), \quad (1)$$

where $m_1, m_2$ and $m_3$ are the masses of three quarks.

To separate the center of mass from the relative coordinates, one can use the Jacobi coordinates [11,14],

$$\vec{R} = \frac{m_1\vec{r_1} + m_2\vec{r_2} + m_3\vec{r_3}}{M}, \quad \vec{\rho} = c_1(\vec{r_1} - \vec{r_2}), \quad \vec{\lambda} = c_2\left(\frac{m_1\vec{r_1} + m_2\vec{r_2}}{M'} - \vec{r_3}\right) \quad (2)$$

Where, $M = m_1 + m_2 + m_3$; $M' = m_1 + m_2$; $c_1$ and $c_2$ are constants.

Using natural units ($\hbar = C = 1$) and letting the coefficients of $\nabla_\rho^2$ and $\nabla_\lambda^2$ be equals to A, equation (1) can be rewritten as;

$$-\frac{1}{2M}\nabla_R^2 \Psi_{C.M}(R) = E_1 \Psi_{C.M}(R) \quad (3)$$

$$\left[-A(\nabla_\rho^2 + \nabla_\lambda^2) + V(x)\right]\Psi(\vec{\rho}, \vec{\lambda}) = E_2 \Psi(\vec{\rho}, \vec{\lambda}) \quad (4)$$

Where, $E = E_1 + E_2$ and

$$A = \frac{c_1^2}{2\mu_{12}} = \frac{c_2^2}{2\mu_{12,3}}.$$

The wave function; $\Psi(\vec{\rho}, \vec{\lambda})$; has been expressed as [8,15,16];

$$\Psi(\vec{\rho},\vec{\lambda}) = \sum_\gamma \Psi_\gamma(x) Y_\gamma(\Omega) \qquad (5)$$

Where, $Y_\gamma(\Omega)$ is called the hyper spherical harmonic function and $\gamma$ is the grand orbital quantum number and takes; 0, 1,2,3,--------------, etc.values.

$$x = \sqrt{\rho^2 + \lambda^2} = \sqrt{\tfrac{1}{3}(r_{12}{}^2 + r_{23}{}^2 + r_{31}^2)},$$

The potential; $V(X)$ ; can be assumed to depend on the hyper radius $x$ ,the space wave function is factorized similarly to the central potential [8] and one term in equation (5) is a solution of equation (4).

Eequation (4) can be rewritten as;

$$\frac{d^2\Psi_\gamma(x)}{dx^2} + \frac{D-1}{x}\frac{d\Psi_\gamma(x)}{dx} - \frac{L^2_{(\Omega)}}{x^2}\Psi_\gamma(x) + \frac{1}{A}(E - V(x))\Psi_\gamma(x) = 0 \qquad (6)$$

Where D represents the dimension of the $\vec{x}$ and $L^2_{(\Omega)}$ is the angular momentum operator whose Eigen functions are [6, 9, 11]:

$$L^2_{(\Omega)} Y_{[\gamma]} = -\gamma(\gamma + D - 2) Y_{[\gamma]} \qquad (7)$$

In the present work, the digamma-function is considered as the interacting potential which is given by;

$$V(x) = C\frac{\Gamma'(x)}{\Gamma(x)} \qquad (8)$$

$$= C \int_0^\infty \left(\frac{e^{-1}}{t} - \frac{e^{-xt}}{1-e^{-t}}\right) dt \qquad (9)$$

Where, C is adjustable parameter and $\Gamma(x)$; is the gamma-function.

Now, applying the following transformations;

$$\Psi_\gamma(x) = u_\gamma(x) / x^{\frac{(D-1)}{2}} \qquad (10)$$

Equation (6) can be rewritten as;

$$\left[\frac{d^2}{dx^2} + \frac{1}{A}\left(E - V(x) - A\left(\frac{(D-1)(D-3)}{4x^2} + \frac{\gamma(\gamma+D-2)}{x^2}\right)\right)\right]u_\gamma(x) = 0 \quad (11)$$

The boundary conditions become:

$$u_\gamma(0) = u_\gamma(\infty) = 0 \quad (12)$$

Using;

$$\lambda = \frac{E}{A} \quad \text{and} \quad \varphi(x) = V(x) + \frac{(D-1)(D-3)}{4x^2} + \frac{\gamma(\gamma+D-2)}{x^2}, \quad (13)$$

equation (10) can be rewritten as;

$$\frac{d^2 u_\gamma(x)}{dx^2} + (\lambda - \varphi(x))u_\gamma(x) = 0 \quad (14)$$

The dimensionless variables g and $\rho(g)$ are given as;

$$g = \frac{1}{1+\frac{x}{x_0}}, \quad \text{and} \quad \rho(g) = g\, u(g), \quad (15)$$

to transform the range of $x \; from \; (\infty, 0) \; to \; (0,1)$.

Then equation (14) can be written as;

$$\frac{d^2\rho(g)}{dg^2} + \left[\frac{x_0^4}{g^4}\right](\lambda - \varphi(g))\rho(g) = 0 \quad (16)$$

The new boundary conditions are given as;

$$\rho(0) = \rho(1) = 0 \quad (17)$$

To transform equation (15) into a matrix form, one can divide the range of (g) from (0,1) into (n+2) points with equal interval; h; labeled by subscript (J).

The new form of the boundary conditions is given as;

$$\rho_{n+1} = \rho_0 \qquad (18)$$

Using the finite difference approximation [17];

$$\frac{d^2\rho\,(g)}{dg^2} = \frac{1}{12h^2}\left[-\rho_{J-2} + 16\rho_{J-1} - 30\rho_J + 16\rho_{J+1} - \rho_{J+2}\right] + O(h^4) \qquad (19)$$

Where, the term $O(h^4)$ represents the expected error.

One can assume that [14],

$$\rho_{-1} = (-1)^\gamma \rho_1 + O(h^2); \qquad (20)$$

$$\rho_{n+2} = (-1)^{\gamma+1} \rho_n + O(h^3); \qquad (21)$$

Using equations (19,20) and 21 into equation (16), and take $x_0 = 1 Gev^{-1}$ one gets;

$$\left(\rho_{J-2} - 16\rho_{J-1} + 30\rho_J - 16\rho_{J+1} + \rho_{J+2}\right) + \frac{12h^2}{(Jh)^4}(\varphi\,(Jh) - \lambda)\rho_J = 0 \qquad (22)$$

This equation is rewritten in the following matrix form;

$$(A - \lambda I)\rho = 0 \qquad (23)$$

Using Jacobi method [18, 19], equation (23) can be solved and the eigen-values; E; are determined. The resonance mass; M; for each state is given by;

$$M = m_1 + m_2 + m_3 + E \qquad (24)$$

## Results and Discussions

The predicted values of the N, Δ, Λ and Σ resonance masses, are calculated through solving equation (23) numerically using the Jacobi method. In this work, the interacting potential between the quark constituents of the considered baryon is represented by the analytical form of digamma-function, see equation (8). The $\chi^2$ −test is used to determine the parameter values;

$$\chi^2 = \frac{1}{N}\sqrt{\sum_{i=1}^{n}\left(\frac{(M_i^{theo}-M_i^{exp})}{e_i}\right)^2} \qquad (25)$$

Where $e_i$ is the experimental error in the $i^{th}$ state.

Table (1) contains the values of the parameters, which are used in the present work for all the considered baryon states. The uncertainties in these parameters may be considered as relativistic effects and the quark flavor dependence of the used potential model.

Table 1: The values of the parameters which are taken in our calculations.

| parameters | Values |
|---|---|
| $m_u$ | 0.035±0.005 $Gev$ |
| $m_d$ | 0.035±0.005 $Gev$ |
| $m_s$ | 0.092±0.008 $Gev$ |
| A | 0.40±0.04 $Gev^{-1}$ |
| C | 0.85±0.08 $Gev$ |

The digamma-function potential versus the hyper radius; x; is shown in fig.(1). In this figure we restrict ourselves to show the behavior of the digamma-function potential at small hyper radius; x; between constituents quarks; $0 < x \leq 2 Gev^{-1}$.

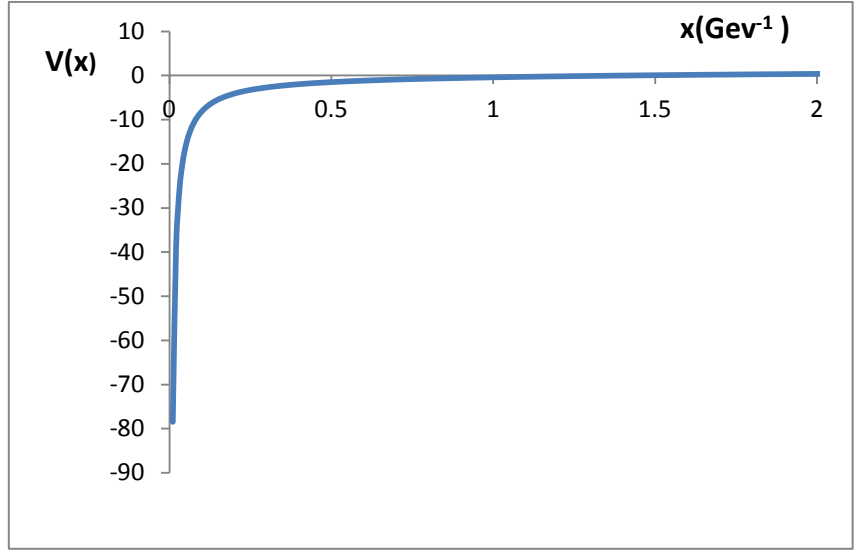

Fig. (1): The digamma –function potential verses the hyper radius x..

Table (2) contains the present calculations of N baryon states using both digamma-function potential and Cornelll potential results [10] in comparison to the experimental values[20]. From this table one notices that, the Cornell potential calculations are 4-states only while in this work, the observed seven experimental states are calculated.  One notices that, the present calculations gives more satisfied agreement with the experiment.

Table 2: The present calculations of N baryon states in comparison to the

Cornell potential calculations[10] and the experimental results in Mev.

| Baryon State | $M_{exp}$ Ref[20] | $\gamma$ | Ref [10] | Present calculations |
|---|---|---|---|---|
| N(938) P11 | 938.272013 ±0.000023 | 0 | 938 | **938.25** |
| N(1440) P11 | 1445±25 | | 1463 | 1448.43 |
| N(1710) P11 | 1710±30 | | 1752 | 1702.24 |
| N(1535) S11 | 1535±10 | 1 | 1524 | 1544.67 |
| N(1905) S11 | 1905 | | - | 1902.34 |
| N(2090) S11 | 2150±50 | | - | 2130.87 |

Table (3) contains the present calculations of Δ baryon states and Cornell potential results [10] in comparison to the allowed experimental values [20].. From this table one can see that, the digamma-function potential results give more agreement with the available experimental values.

Table 3: The present calculations of Δ baryon states in comparison to the Cornell potential calculations[10] and the experimental results in (MeV.).

| Baryon State | $M_{exp}$ Ref [20] | $\gamma$ | Ref [10] | Present calculations |
|---|---|---|---|---|
| Δ (1232) P33 | 1232±1 | 0 | 1232 | 1233.24 |
| Δ (1600) P33 | 1625±75 | | 1727 | 1594.57 |
| Δ (1620) S31 | 1630±30 | 1 | 1573 | 1612.95 |
| Δ (1900) S31 | 1900 | | - | 1896.32 |
| Δ (2150) S31 | 2150 | | - | 2160.73 |

The present calculations of Λ and Σ baryon states compared with the experimental values [20] are given in tables (4-5). One notices that; the digamma-function results give a satisfied agreement with the experiment.

Table 4: Present calculations of Λ baryon states in comparison to the experimental results in (MeV).

| Baryon State | $M_{exp}$ Ref [20] | $\gamma$ | Present calculations |
|---|---|---|---|
| Λ(1116) P01 | 1115.683±0.006 | 0 | 1116.85 |
| Λ(1600) P01 | 1630±70 | | 1689.46 |
| Λ(1810) P01 | 1800±50 | | 1832.86 |
| Λ(1670) S01 | 1670±10 | 1 | 1665.21 |
| Λ(1800) S01 | 1785±65 | | 1851.75 |
| Λ(1890) P03 | 1880±30 | 2 | 1915.79 |

Table 5: The present calculations of Σ baryon states in comparison to the experimental results in (MeV.)

| Baryon State | $M_{exp}$ Ref. [20] | $\gamma$ | Present calculations |
|---|---|---|---|
| Σ(1193) P11 | 1192.642±0.024 | 0 | 1193.15 |
| Σ(1660) P11 | 1660±30 | 0 | 1629.47 |
| Σ(1880) P11 | 1880 | 0 | 1868.13 |
| Σ(1620) S11 | 1620 | 1 | 1617.39 |
| Σ(1750) S11 | 1765±35 | 1 | 1809.35 |

In case of N and Δ baryons, one can notice that, the calculated resonance masses using the digamma-function potential results more agree with the experimental values than the Cornelll potential results[10]. Also our calculations using the present potential model can reproduce the experimental results of Λ and Σ baryon states. Finally, one can conclude that; the non-relativistic study of the baryon systems; as three body problem; using digamma -function as interacting potential between its constituent quarks; may be suitable in studding the excited states of different baryon systems.